\documentclass[twocolumn]{el-author}

\newcommand{\ua}{\uparrow}
\newcommand{\nc}{\newcommand}

\usepackage{algorithm}
\usepackage{algorithmic}
\nc{\da}{\downarrow} \nc{\hc}{\hat{c}} \nc{\hS}{\hat{S}}
\nc{\bra}{\langle} \nc{\ket}{\rangle} \nc{\eq}{equation (\ref}
\nc{\h}{\hat} \nc{\hT}{\h{T}}\nc{\be}{\begin{eqnarray}}
\nc{\ee}{\end{eqnarray}}\nc{\rd}{\textrm{d}}\nc{\e}{eqnarray}\nc{\hR}{\hat{R}}\nc{\Tr}{\mathrm{Tr}}
\nc{\tS}{\tilde{S}}\nc{\tr}{\mathrm{tr}}\nc{\8}{\infty}\nc{\lgs}{\bra\ua,\phi|}\nc{\rgs}{|\ua,\phi\ket}
\nc{\hU}{\hat{U}}\nc{\lfs}{\bra\phi|}\nc{\rfs}{|\phi\ket}\nc{\hZ}{\hat{Z}}\nc{\hd}{\hat{d}}\nc{\mD}{\mathcal{D}}
\nc{\bd}{\bar{d}}\nc{\bc}{\bar{c}}\nc{\mc}{\mathcal}\nc{\ea}{eqnarray}\nc{\mG}{\mathcal{G}}\nc{\bce}{\begin{center}}
\nc{\ece}{\end{center}}
\date{5th March 2014}

\begin{document}

\title{Structured Compressive Sensing Based Superimposed Pilot Design in Downlink Large-Scale MIMO Systems}

\author{Zhen Gao, Linglong Dai and Zhaocheng Wang, \emph{IET Fellow}}

\abstract
{Large-scale multiple-input multiple-output (MIMO) with high spectrum and energy efficiency is a very promising key technology for future 5G wireless communications. For large-scale MIMO systems, accurate channel state information (CSI) acquisition is a challenging problem, especially when each user has to distinguish and estimate numerous channels coming from a large number of transmit antennas in the downlink. Unlike the conventional orthogonal pilots whose pilot overhead prohibitively increases with the number of transmit antennas, we propose a spectrum-efficient superimposed pilot design for downlink large-scale MIMO scenarios, where frequency-domain pilots of different transmit antennas occupy the completely same subcarriers in the freqency domain. Meanwhile, spatial-temporal common sparsity of large-scale MIMO channels motivates us to exploit the emerging theory of structured compressive sensing (CS) for reliable MIMO channel estimation, which is realized by the proposed structured subspace pursuit (SSP) algorithm to simultaneously recover multiple channels with low pilot overhead. Simulation results demonstrate that the proposed scheme performs well and can approach the performance bound.
}

\maketitle
\vspace*{-0.5mm}
\section{Introduction}
As a promising key technology for future 5G wireless communications, large-scale multiple-input multiple-output (MIMO) employing a large number of antennas at the base stations (BS) to simultaneously serve multiple users can boost the spectrum efficiency and energy efficiency by orders of magnitude \cite{1}. In large-scale MIMO systems, accurate channel state information (CSI) is essential for signal detection, precoding, resource allocation, etc., and its reliable acquisition is challenging, especially in downlink where each user has to accurately estimate channels from large numbers of BS antennas \cite{2}. By far, most researches on large-scale MIMO assume time division duplexing (TDD) protocol, where the acquired CSI at the BS in uplink can be directly feedback to users due to the channel reciprocity property, thus the challenging CSI acquisition in downlink can be avoided \cite{3}. However, the CSI obtained in uplink may not be accurate or even outdated for the downlink in TDD systems, which will cause a significant performance loss, especially for mobile users. Moreover, since frequency division duplexing (FDD) still dominates current wireless cellular systems, the downlink CSI must be acquired since the channel reciprocity property does not exist for FDD systems \cite{FDD}.

In this letter, we propose a spectrum-efficient superimposed pilot design based on the emerging theory of structured compressive sensing (CS) \cite{STR_CS} to solve the challenging problem of accurate CSI acquisition in downlink large-scale MIMO systems. In contrast to the conventional orthogonal pilots whose pilot overhead prohibitively increases with the number of transmit antennas due to the pilots for different transmit antennas are orthogonal in the frequency domain, the proposed superimposed pilot design allow different transmit antennas to occupy the exactly same subcarriers. At the receiver, the structured subspace pursuit (SSP) algorithm derived from the classical SP algorithm \cite{SP} in standard CS theory is proposed to reliably distinguish and accurately estimate multiple channels by exploiting the inherent spatial-temporal common sparsity of large-scale MIMO channels \cite{SCS}. In this way, the pilot overhead can be significantly reduced.
\vspace*{-0.1mm}
\section{Spatial-Temporal Common Sparsity of Large-Scale MIMO Channels}
During the $k$th OFDM symbol, the channel impulse response (CIR) between the $m$th transmit antenna (the total number of transmit antennas is $M$) and a certain user can be denoted by ${{\bf{h}}_{m,k}} = {\left[ {{h_{m,k}}(0),{h_{m,k}}(1), \cdots ,{h_{m,k}}(L - 1)} \right]^T}$ for $1 \le m \le M$, where $L$ is the maximum channel delay spread, and the number of nonzero elements $K$ in ${{\bf{h}}_{m,k}}$ satisfies $K \ll L$ due to the sparsity of wireless channels \cite{SCS}. Meanwhile, due to the compact antenna geometry, CIRs between different transmit-receive pairs share very similar path delays, which is referred as the spatial common sparsity of MIMO channels \cite{SCS}. Thus, let $S_{m,k}  = \text{supp}\{{{\bf{h}}_{m,k}}\}=\left\{ {\tau :\left| {{h_{m,k}}(\tau )} \right| > 0} \right\}_{\tau  = 0}^{L - 1}$ be the support of ${{\bf{h}}_{m,k}}$, we have $S_{1,k}  = S_{2,k}  =  \cdots  = S_{M,k} $.

Moreover, during several adjacent OFDM symbols, although path gains may be quite different, path delays remain almost unchanged \cite{Dai}. Hence MIMO channels also exhibit temporal common sparsity. Therefore, we further have $S_{m,k}  = S_{m,k + 1}  =  \cdots  = S_{m,k + R - 1}$, and the temporal common sparsity can be assumed during $R$ adjacent OFDM symbols \cite{Dai}.

Such spatial-temporal common sparsity of MIMO channels, which is usually not considered by state-of-the-arts schemes, will be exploited for the spectrum-efficient pilot design and reliable CSI acquisition in this letter.

\begin{figure}[!tp]
     \centering
     \includegraphics[width=4.7cm, keepaspectratio,angle=90]
     {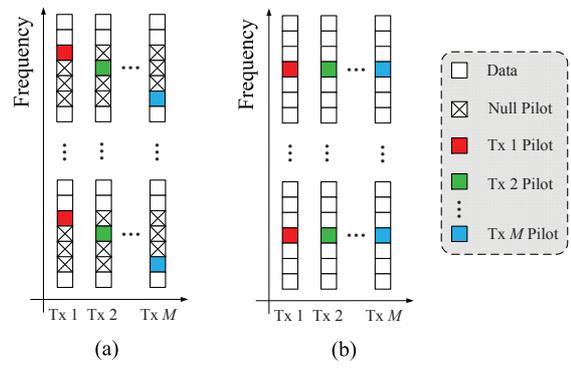}
         \vspace*{-2mm}
    \caption{Comparison of the conventional orthogonal pilots and the proposed superimposed pilots. (a): Conventional orthogonal pilots; (b): Proposed superimposed pilots.}
         \vspace*{-4.80mm}
     \label{fig:pilot}
\end{figure}

\section{Superimposed Pilot Design}
In the conventional orthogonal pilots as shown in Fig. \ref{fig:pilot} (a), pilots of different transmit antennas are non-overlapping, and null pilots are used to prevent the mutual interferences among different transmit antennas at the cost of high pilot overhead, especially when the number of antennas becomes large.
On the contrary,  the proposed superimposed pilot design allows pilots of different transmit antennas to occupy the completely same subcarriers in the frequency domain as illustrated in Fig. \ref{fig:pilot} (b), and the large pilot overhead due to null pilots can be avoided.
Specifically, subcarrier indexes allocated to pilots can be denoted by $\Omega$, which is identical for all transmit antennas, and it can be uniformly decimated from $\left[ {0,N - 1} \right]$ as the conventional comb-type pilots for single-antenna systems, where $N$ is the discrete Fourier transform (DFT) size of the OFDM symbol. Meanwhile, in order to distinguish channels associated with different transmit antennas, pilot sequences $\{ {{\bf{p}}_m}\} _{m = 1}^M$ of different transmit antennas differ one from another, i.e., ${{\bf{p}}_m} \ne {{\bf{p}}_n}$ if $m \ne n$, which can be easily realized by generating the pilot sequences according to the independent and identically distributed (i.i.d.) random Bernoulli distribution ($ \pm 1$). Consequently, in contrast to the conventional orthogonal pilots with the total pilot overhead ${N_{p{\rm{\_total}}}} = M{N_p}$, where $N_p$ denotes the required number of pilots per transmit antenna, the total pilot overhead in the proposed scheme is reduced to ${N_{p{\rm{\_total}}}} = {N_p}$ due to the superimposed pilot design.

At the receiver, after the removal of the cyclic prefix and DFT, the received pilot sequence ${\bf{y}}_k$ of the $k$th OFDM symbol coming from $M$ different transmit antennas can be expressed as
\begin{equation}\label{equ:re_pilot}
\begin{array}{l}
{{\bf{y}}_k} = \sum\limits_{m = 1}^M {{\rm{diag}}\{ {{\bf{p}}_{m}}\} {{\left. {\bf{F}} \right|}_\Omega }\left[ {\begin{array}{*{20}{l}}
{{{\bf{h}}_{m,k}}}\\
{{{\bf{0}}_{(N - L) \times 1}}}
\end{array}} \right]}  + {\bf{w}}_{k}\\
 {\kern 10pt} = \sum\limits_{m = 1}^M {{{\bf{P}}_m}{{\left. {{{\bf{F}}_L}} \right|}_\Omega }{{\bf{h}}_{m,k}}}  + {\bf{w}}_{k},
\end{array}
\end{equation}
where ${{\bf{P}}_m} = {\rm{diag}}\{ {{\bf{p}}_m}\}$ is a diagonal matrix with ${{\bf{p}}_m}$ on its diagonal, ${\bf{F}}$ is a DFT matrix of size $N \times N$, ${{\bf{F}}_L}$ is a partial DFT matrix of size $N \times L$ consisted of the first $L$ columns of ${\bf{F}}$, ${\left. {{{\bf{F}}_L}} \right|_\Omega }$ denotes the sub-matrix by selecting the rows of ${{{\bf{F}}_L}}$ according to $\Omega $, and ${\bf{w}}_k$ is the additive white Gaussian noise (AWGN). Moreover, (\ref{equ:re_pilot}) can be also rewritten in a more compact form as
\begin{equation}\label{equ:compact}
{\bf{y}}_k = {\bf{\Phi {\tilde h}}}_k + {\bf{w}}_k,
\end{equation}
where ${\bf{\Phi }} = \left[ {{{\bf{P}}_1}{{\left. {{{\bf{F}}_L}} \right|}_\Omega },{{\bf{P}}_2}{{\left. {{{\bf{F}}_L}} \right|}_\Omega }, \cdots ,{{\bf{P}}_M}{{\left. {{{\bf{F}}_L}} \right|}_\Omega }} \right]$ has the size of ${N_p} \times ML$, and ${\bf{\tilde h}}_k = {[{\bf{h}}_{1,k}^T,{\bf{h}}_{2,k}^T, \cdots ,{\bf{h}}_{M,k}^T]^T}$ is an equivalent CIR vector of size $ML \times 1$.

For large-scale MIMO systems, we usually have ${N_p} \ll ML$ due to numerous transmit antennas $M$ and the limited number of pilots $N_p$, which implies that we cannot reliably recover ${\bf{\tilde h}}_k$ from ${\bf{y}}_k$ according to the underdetermined problem (\ref{equ:compact}). However, the observation that ${\bf{\tilde h}}_k$ is a sparse signal due to the sparsity of $\{ {{\bf{h}}_{m,k}}\} _{m = 1}^M$ inspires us to reconstruct the high-dimension sparse signal ${\bf{\tilde h}}_k$ from the low-dimension received pilot sequence ${\bf{y}}_k$ under the framework of CS theory. Moreover, the spatial-temporal common sparsity of wireless MIMO channels can be also integrated in the classical CS framework for performance enhancement, which is the topic of the following section.

\section{Joint Sparse Channel Estimation Based on Structured CS}

\vspace*{+2.00mm}
\begin{algorithm}[tp]
\renewcommand{\algorithmicrequire}{\textbf{Input:}}
\renewcommand\algorithmicensure {\textbf{Output:} }
\caption{ Proposed  Structured Subspace Pursuit (SSP) Algorithm.}
\label{alg:Framwork} 
\begin{algorithmic}[1]
\REQUIRE
Noisy measurement matrix ${\bf{Y}}$ and sensing matrix ${\bf{\Phi }}$ in (\ref{equ:common}).
\ENSURE
Estimated CIR matrix $\widehat {\bf{H}}$. \\

\STATE $\Omega  \leftarrow \emptyset$;
\STATE $k \leftarrow 1$;
\STATE ${{\bf{V}}_1} \leftarrow {\bf{Y}}$;

\WHILE{${\left\| {{{{\bf{V}}}_k}} \right\|_F}  <  {\left\| {{{\bf{V}}_{k - 1}}} \right\|_F}$,}

\STATE ${\bf{Z}} \leftarrow {{\bf{\Phi }}^H}{{\bf{V}}_k}$;
\STATE ${c}(\tau ) \leftarrow \sum\nolimits_{r = 1,i = 0}^{R,M - 1} {{{\left| {{z^{(\tau  + iL,r)}}} \right|}^2}}$, $0 \le \tau  \le L - 1$;
\STATE $\Omega  \leftarrow \Omega  \cup {\rm{supp}}\{{\left. {\bf{c}} \right\rangle _K}\} $;
\STATE $\Gamma  \leftarrow \Omega  \cup \left[ {\Omega  + L} \right] \cup  \cdots  \cup \left[ {\Omega  + L(M - 1)} \right]$;
\STATE ${\left. {{\bf{\hat H}}} \right|_\Gamma } \leftarrow {\bf{\Phi }}_\Gamma ^\dag {\bf{Y}}$;
\STATE ${r}(\tau ) \leftarrow \sum\nolimits_{r = 1,i = 0}^{R,M - 1} {{{\left| {{{\hat h}^{(\tau  + iL,r)}}} \right|}^2}}$, $0 \le \tau  \le L - 1$;
\STATE $\Omega  \leftarrow  {\rm{supp}}\{{\left. {\bf{r}} \right\rangle _K}\} $;
\STATE $\Gamma  \leftarrow \Omega  \cup \left[ {\Omega  + L} \right] \cup  \cdots  \cup \left[ {\Omega  + L(M - 1)} \right]$;
\STATE ${\left. {{\bf{\hat H}}} \right|_\Gamma } \leftarrow {\bf{\Phi }}_\Gamma ^\dag {\bf{Y}}$;
\STATE ${{\bf{V}}_k} \leftarrow {\bf{Y}} - {{\bf{\Phi }}^H}\widehat {\bf{H}}$;
\STATE $k \leftarrow k + 1$;
\ENDWHILE
\end{algorithmic}
\end{algorithm}
\vspace*{-6.00mm}

The spatial-temporal common sparsity of wireless MIMO channels as addressed above motivates us to exploit the theory of structured CS  \cite{STR_CS} developed from the classical CS theory to simultaneously reconstruct multiple channels. Considering (\ref{equ:compact}) during $R$ adjacent OFDM symbols with the same pilot pattern, which is quite common in practice \cite{SCS}, we have
\begin{equation}\label{equ:common}
{\bf{Y}} = {\bf{\Phi H}} + {\bf{W}},
\end{equation}
where ${\bf{Y}} = \left[ {{{\bf{y}}_k},{{\bf{y}}_{k + 1}}, \cdots ,{{\bf{y}}_{k + R - 1}}} \right]$, ${\bf{H}} = \left[ {{{{\bf{\tilde h}}}_k},{{{\bf{\tilde h}}}_{k + 1}}, \cdots ,{{{\bf{\tilde h}}}_{k + R - 1}}} \right]$, and ${\bf{W}} = \left[ {{{\bf{w}}_k},{{\bf{w}}_{k + 1}}, \cdots ,{{\bf{w}}_{k + R - 1}}} \right]$.

According to the property of the spatial-temporal common sparsity in MIMO channels, vectors consisting of ${\bf{H}}$ share the common sparse support. Therefore, ${\bf{H}}$ has the inherent structured sparsity. Based on the classical subspace pursuit (SP) algorithm, which reconstructs a single high-dimension sparse vector from one low-dimension noisy measurement vector \cite{Dai}, we propose the structured SP (SSP) algorithm which can jointly recover multiple high-dimension sparse vectors with structured sparsity from multiple low-dimension noisy measurement vectors. The proposed SSP algorithm is described in Algorithm 1. Note that in Algorithm 1, ${\left\| \cdot\ \right\|_F}$ denotes the $F$-norm operation, $(\cdot )^{ H}$ denotes the conjugate transpose, $(\cdot )^{\dag}$ denotes the Moore-Penrose matrix inversion, ${{\bf{c}}} = {\left[ {{c}(0),{c}(1), \cdots ,{c}(L - 1)} \right]^T}$, ${{\bf{r}}} = {\left[ {{r}(0),{r}(1), \cdots ,{r}(L - 1)} \right]^T}$, $\left[ {\Omega  + a} \right]$ means to add $a$ to each element of $\Omega$, the vector ${\left. {\bf{x}} \right\rangle _K}$ is generated by retaining the $K$ largest elements of $\bf {x}$ while setting the rest elements to zero, ${{\bf{\Phi }}_\Gamma }$ denotes the sub-matrix by selecting the columns of ${\bf{\Phi }}$ according to $\Gamma$, $z^{(i,j)}$ and ${\hat h}^{(i,j)}$ denote the $i$th row and the $j$th column elements of $\bf{Z}$ and $\bf{\hat H}$, respectively.

Compared with the classical SP algorithm where only one vector is updated in each iteration, the proposed SSP algorithm simultaneously updates multiple vectors in each iteration since taps with the same location in each CIR vectors are considered as a whole.
\section{Simulation Results}
\begin{figure}[!tp]
     \centering
     \includegraphics[width=8.3cm, keepaspectratio]
     {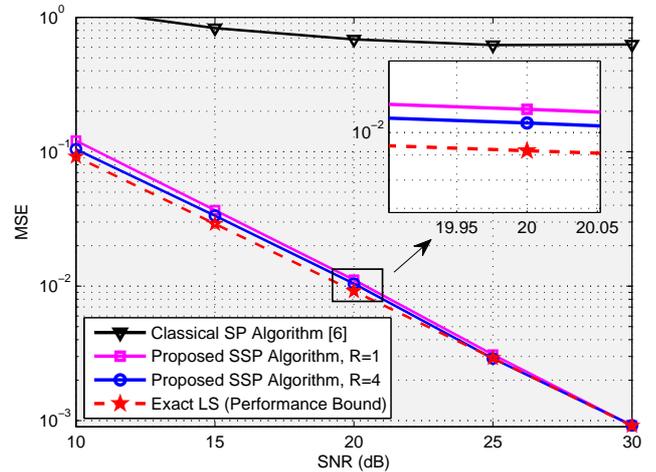}
         \vspace*{-1.50mm}
    \caption{MSE performance comparison over the ITU Vehicular B channel.}
         \vspace*{-2.00mm}
     \label{fig:MSE}
\end{figure}

A simulation study was carried out to investigate the mean square error (MSE) performance of the proposed solution. The proposed SSP algorithm with different $R$'s, the classical SP algorithm, and the exact least square (LS) method with perfectly known support of the sparse channel as the performance bound were compared based on the proposed superimposed pilot scheme. System parameters were set as follows: the number of transmit antennas is $M=64$, the system bandwidth is 10 MHz, the DFT size of the OFDM symbol is $N = 4096$, the cyclic prefix length is $N_g = 256$, and $N_p = 800$ superimposed pilots are uniformly spaced in the frequency domain. Besides, the six-path International Telecommunication Union (ITU) Vehicular B channel model \cite{Dai} with $L = 200$ was considered. Fig. \ref{fig:MSE} indicates that the classical SP algorithm cannot work since ${N_p}$ is too small to reliably recover a ($MK$)-sparse signal of size $ML$ with $N_p \ll ML$. On the contrary, the proposed SSP algorithm with $R = 4$ works better than that with $R = 1$, and both of them perform well and approach to the exact LS method. For the proposed superimposed pilot design, $N_{p\_\text{total}} = 800$ pilots only occupy $19.53\%$ of the total $N=4096$ subcarriers, and the equivalent average pilot overhead per antenna is just $N_{p\_\text{avg}} = 12.5$ (only $0.31\%$ of the total $N=4096$ subcarriers). This implies that compared with the classical CS algorithms (e.g., SP algorithm) which usually require $N_{p\_\text{avg}}={\cal{O}}(K\text{log}_2N)$ pilots, $N_{p\_\text{avg}}$ required by the proposed SSP algorithm approaches to the theoretical limit of $2K=12$ \cite{STR_CS}. Meanwhile, such low pilot overhead is almost impossible for the conventional channel estimation schemes to realize accurate channel estimation.

It is worth noting that although $\Omega $ and $\{ {{\bf{p}}_m}\} _{m = 1}^M$ used in this letter guarantee the near-orthogonal columns of the sensing matrix ${\bf{\Phi }}$ and thus the reliable performance, the optimal superimposed pilot design remains an interesting problem to be studied in the future.
\section{Conclusions}
This letter investigates the challenging problem of pilot design and channel estimation for downlink large-scale MIMO systems. Compared with conventional orthogonal pilots which suffer from the prohibitive overhead in large-scale MIMO systems, the proposed superimposed pilot design based on structured CS can efficiently reduce the pilot overhead. Meanwhile, the proposed SSP algorithm exploiting the spatial-temporal common sparsity of large-scale MIMO channels can accurately recover multiple channels simultaneously. Moreover, the proposed superimposed pilot design and the structured CS based channel estimator can be also extended to the uplink as well as conventional small-scale MIMO systems for accurate CSI acquisition with low pilot overhead.

\vspace*{-2.50mm}
\vskip3pt \ack{This work was supported by National Key Basic
Research Program of China (Grant No. 2013CB329201), National Natural
Science Foundation of China (Grant No. 61201185), and the ZTE fund
project (Grant No. CON1307250001). }

\vskip5pt

\noindent Zhen Gao, Linglong Dai and Zhaocheng
 Wang (\textit{Tsinghua National
Laboratory for Information Science and Technology, Department of
Electronic Engineering, Tsinghua University, Beijing 100084, China})

\vskip3pt

\noindent E-mail: daill@tsinghua.edu.cn

\end{document}